# High-order Order Proximity-Incorporated, Symmetry and Graph-Regularized Nonnegative Matrix Factorization for Community Detection

Zhigang Liu, and Xin Luo, *Senior Member, IEEE*

*Abstract*—Community describes the functional mechanism of a network, making community detection serve as a fundamental graph tool for various real applications like discovery of social circle. To date, a Symmetric and Non-negative Matrix Factorization (SNMF) model has been frequently adopted to address this issue owing to its high interpretability and scalability. However, most existing SNMF-based community detection methods neglect the high-order connection patterns in a network. Motivated by this discovery, in this paper, we propose a High-Order Proximity (HOP)-incorporated, Symmetry and Graph-regularized NMF (HSGN) model that adopts the following three-fold ideas: a) adopting a weighted pointwise mutual information (PMI)-based approach to measure the HOP indices among nodes in a network; b) leveraging an iterative reconstruction scheme to encode the captured HOP into the network; and c) introducing a symmetry and graph-regularized NMF algorithm to detect communities accurately. Extensive empirical studies on eight real-world networks demonstrate that an HSGN-based community detector significantly outperforms both benchmark and state-of-the-art community detectors in providing highly-accurate community detection results.

*Index Terms*—Undirected Network, Network Representation Learning, Community Detection, Non-negative Matrix Factorization, Symmetry and Graph regularization.

## I. Introduction

A NETWORK can be taken as an abstraction of a complex application system in real life, e.g., a social networking site [1], and an IoT-based smart system [2], which involves numerous entities and their interactions. A prominent feature of a network is the underlying community structures [3], which play a significant role in describing the functional mechanism of the network. Community reveals partial network nodes with more frequent interactions, which are particularly helpful in understanding the networking organization mechanism and extracting useful knowledge form a network [4, 8, 9].

A Symmetric and Nonnegative Matrix Factorization (SNMF) model [4, 10] possesses good interpretability, high scalability and efficiency, and good compatibility with clustering-related tasks. It has been frequently adopted to develop a community detector on undirected networks which are the most common form in real applications [5, 6, 10]. However, most existing SNMF-based detectors [4, 8, 10, 16] are built depending on the first-order proximities that is inadequate to precisely represent the network [7]. To leverage the information beyond the first-order network topology, existing pioneering studies mainly have carried out the following two-fold attempts.

First, by adopting some additional prior information such as pairwise constraints [11, 12] and node attributes [13-15] as the supplement data to learn the final network representation, semi-supervised community detectors yielded. For instance, Lu *et al.* [11] and Shi *et al.* [12] encode the pairwise constraints on nodes' prior group information into graph-regularization to boost an SNMF-based model's community detection accuracy. Hong *et al.* [14] propose a deep attributed network embedding model to capture the complex structure and attribute information, which adopts a step-based random walk to capture the interaction between network structure and node attributes from various degrees of proximity. Commonly, with effective supplementary additional information, those methods achieve excellent community detection accuracy. However, their performance of a semi-supervised community detector relies heavily on the quality of prior information, which is not always met in practice.

Second, some efforts [19-22] have been made to take into account both the first-order and second-order node proximities together to describe network topology, thereby improving a detector's performance. For instance, Tang *et al.* [19] propose a Large-scale Information Network Embedding (LINE) model to capture the second-order node similarity by considering node pairs with common neighbors. Ye *et al.* [21] propose a homophily preserving NMF model that combines both the link topology and node homophily (i.e., the node proximity revealed by specific similarity measurements) of a network, thereby better describing community structures. To handle the first- and second-order proximity information efficiently, Luo *et al.* [22] utilize Pointwise Mutual Information (PMI) to quantify the second-order node proximity, and propose a PMI-incorporated Graph-regularized SNMF (PGS) model for highly-accurate community detection. However, these methods still have some limitations, e.g., they rarely take high-order node interactions into consideration.

Besides, existing NMF-based community detectors mostly adopt a standard SNMF model to perform community detection, which adopts one unique LF matrix only to learn the network representation for ensuring the rigorous symmetry. With such design, the representation learning ability of a resultant model

Z. Liu and X. Luo are with the School of Computer Science and Technology, Chongqing University of Posts and Telecommunications, Chongqing 400065, China, and with the Chongqing Institute of Green and Intelligent Technology, Chinese Academy of Sciences, Chongqing 400714, China (email: liuzhigangx@gmail.com, luoxin21@gmail.com).

is reduced significantly, which inevitably affects its community detection accuracy [23-27].

Aiming at addressing the above issues, this paper proposes a High-Order Proximity (HOP)-incorporated, Symmetry and Graph-regularized Nonnegative Matrix Factorization (HSGN) model relying on three-fold ideas: a) proposing a weighted PMI-based approach to quantize the HOP indices among nodes based on the high-order neighbor node pairs extracted from a network; b) leveraging an iterative reconstruction scheme to rebuild some crucial but missing links in the network based on acquired HOP information; and c) introducing a Symmetry and Graph-regularized NMF (SGN) algorithm for community detection.

This paper makes the following contributions:

a) **Proposing an HOP-incorporated iterative network reconstruction scheme.** Based on the discovery of the HOP' positive effects in enhancing the target network, this paper considers multiple proximities among nodes. An iterative reconstruction process is performed to propagate the effects of HOP to enhance the network topology.

b) **Implementing an HSGN-based community detector.** With an SGN algorithm, an HSGN-based community detector represents an undirected network by leveraging symmetry-regularizations that imply equality constraints on its multiple LF matrices to maintain symmetry while guaranteeing the model capacity for enhancing its representation ability. With the HOP-enhanced network as the fundamental input, its input information is significantly enhanced to enable its highly-accurate detection results.

Empirical studies on ten networks from real applications demonstrate that comparing with state-of-the-art models, the proposed HSGN model achieves significant accuracy gain for community detection.

Section II gives the preliminaries. Section III presents an HOP-NMF-based community detector. Extensively empirical studies and analyses are provided in Section IV. Finally, conclusions are drawn in Section V.

## II. PRELIMINARIES

### A. Problem Statement

This paper considers an undirected and unweighted network $G=(V, E)$, where $V=\{v_1, v_2, …, v_n\}$ denotes a set of $n$ nodes and $E=\{e_1, e_2, …, e_m\}$ denotes a set of $m$ edges. The topology of $G$ is described with its adjacency matrix $A \in \mathbb{R}^{n \times n}$. Note that $A$ is a symmetric, nonnegative and binary matrix, whose $(i, j)$-th entry $a_{ij}$ is designated as one if there is an edge between nodes $v_i$ and $v_j$, and zero otherwise. Since it preserves directly connected edges among nodes in a target network, the adjacency matrix actually describes the first-order proximity (FOP) of the target network.

Given $G$ whose community count $K$ is given in prior, a community detector aims to find a community set as $C=\{C_i | C_i \neq \emptyset, \cup_{i=1}^{K} C_i = V, C_i \neq C_j, 1 \leq i \leq K, 1 \leq j \leq K\}$, where $C_i$ denotes the $i$-th community in $C$, and $\cup_{i=1}^{K}$ calculates the union set.

### B. Pointwise Mutual Information

PMI is originally designed to measure the association of two words in natural language processing [28]. It compares the probability of observing two entities simultaneously with that of observing them independently:

$$I_{PMI}(w_1, w_2) = \log \frac{p(w_1, w_2)}{p(w_1) p(w_2)}, \quad (1)$$

where $p(w_1)$ and $p(w_2)$ calculate the probabilities of observing entities $w_1$ and $w_2$ independently while $p(w_1, w_2)$ calculates the probability of observing $w_1$ and $w_2$ simultaneously. Note that (1) indicates if $p(w_1, w_2)$ is larger than $p(w_1)p(w_2)$, then an association relationship between $w_1$ and $w_2$ should be considered. In this work, we want to use PMI as a useful metric to quantize high-order connection relationships among nodes, thereby capturing their HOP information.

### C. Community Detection via SNMF

Given a network $G$ with $A$, an SNMF-based community detector aims to learn an approximation $\hat{A}$ to $A$ with a unique LF matrix $X^{n \times K}$ to achieve $\hat{A}=XX^T$. While $X$ represents information in $A$, it can be considered as the soft indicator demonstrating the node-community affiliation as its each entry $x_{il}$ demonstrates the probability that $v_i \in C_l$.

To make $X$ precisely represent $A$, a loss function depending on Euclidean distance is then formulated as [29]:

$$J_{SNMF} = \frac{1}{2} \left\| A - XX^T \right\|_F^2, \quad s.t. X \geq 0, \quad (2)$$

where $\|\cdot\|$ calculates the Frobenius norm of an enclosed matrix.

By considering the Karush-Kuhn-Tucker (KKT) conditions of (2) [23, 29], the following nonnegative multiplicative update (NMU) learning scheme of $X$ is achieved:

$$x_{ik} \leftarrow x_{ik} \left( (AX)_{ik} \big/ (XX^T X)_{ik} \right). \quad (3)$$

An SNMF model is obtained by the following commonly adopted learning rule with an adjusted multiplicative to ensure its stable convergence [29]:

$$x_{ik} \leftarrow x_{ik} \left( 0.5 + \left( (AX)_{ik} \big/ (2XX^T X)_{ik} \right) \right). \quad (4)$$

With $X$, an SNMF-based community detector interprets it as a soft community-membership indicator, i.e., $\forall i \in \{1, 2, …, n\}$ and $l \in \{1, 2, …, K\}$, $x_{il}$ describes the probability that node $v_i$ belongs to community $C_l$. Such a process is formulated as:

$$\forall v_i \in V: v_i \in C_l, \text{ if } x_{il} = \max\{x_{ik} | k \in \{1, 2, …, K\}\}. \quad (5)$$

Note that division mechanism (5) is a widely adopted by most of NMF-based community detection methods. It is also applicable to this paper.

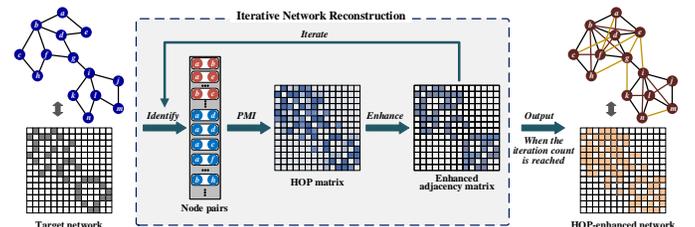

Fig. 1. An illustration of the proposed HOP-incorporated iterative reconstruction scheme.

## III. AN HSGN-BASED COMMUNITY DETECTION MODEL

### A. A Novel HOP Metric

To quantify the HOP indices among nodes, we propose a weighted PMI-based metric. Given an arbitrary node pair $\{v_i, v_j\}$, we calculate the probabilities of observing them as:

$$p(v_i) = \frac{N(v_i)}{|H|} = \omega_1 \frac{N_1(v_i)}{|H_1|} + \omega_2 \frac{N_2(v_i)}{|H_2|} + \cdots + \omega_n \frac{N_r(v_i)}{|H_r|} \quad (6a)$$
$$= \sum_{k=1}^{r} \left( \omega_k N_k(v_i) / |H_k| \right),$$

$$p(v_j) = \frac{N(v_j)}{|H|} = \omega_1 \frac{N_1(v_j)}{|H_1|} + \omega_2 \frac{N_2(v_j)}{|H_2|} + \cdots + \omega_n \frac{N_r(v_j)}{|H_r|} \quad (6b)$$
$$= \sum_{k=1}^{r} \left( \omega_k N_k(v_j) / |H_k| \right),$$

$$p(v_i, v_j) = N(v_i, v_j) / |H|$$
$$= \omega_1 \frac{N_1(v_i, v_j)}{|H_1|} + \omega_2 \frac{N_2(v_i, v_j)}{|H_2|} + \cdots + \omega_n \frac{N_r(v_i, v_j)}{|H_r|} \quad (6c)$$
$$= \sum_{k=1}^{r} \left( \omega_k N_k(v_i, v_j) / |H_k| \right),$$

where $v_i$ and $v_j$ denote two observed nodes. $p(v_i)$ and $p(v_j)$ denote the probabilities of observing all of the node pairs that contain $v_i$ and $v_j$, and $p(v_i, v_j)$ is the probability of observing the node pair $\{v_i, v_j\}$; $N_k(v_i)$ and $N_k(v_j)$ denote the number of the $k$-th-order node pairs that contains node $v_i$ and $v_j$; $N_k(v_i, v_j)$ denotes the appearance count of the $k$-th-order node pair $\{v_i, v_j\}$, $H_k$ denotes a set of $k$-th-order node pairs, and $|H_k|$ denotes the number of node pairs in $H_k$. $\omega_k$ is the weight for the $k$th-order node proximity.

With (6), we calculate the HOP index between $v_i$ and $v_j$ as:

$$P(v_i, v_j) = \log \left( p(v_i, v_j) / p(v_i) p(v_j) \right)$$
$$= \log \left( \sum_{k=1}^{r} \left( \omega_k \frac{N_k(v_i, v_j)}{|H_k|} \right) \bigg/ \sum_{k=1}^{r} \left( \omega_k \frac{N_k(v_i)}{|H_k|} \right) \cdot \sum_{k=1}^{r} \left( \omega_k \frac{N_k(v_j)}{|H_k|} \right) \right). \quad (7)$$

Intuitively, the importance of HOP will decay with the increase of a node pair's order that has been taken into consideration. From this point, we take the reciprocal of the proximity as the weighting coefficient for HOP indices, thereby extending (7) into:

$$P(v_i, v_j) = \log \left( \sum_{k=1}^{r} \left( \frac{N_k(v_i, v_j)}{k|H_k|} \right) \bigg/ \sum_{k=1}^{r} \left( \frac{N_k(v_i)}{k|H_k|} \right) \cdot \sum_{k=1}^{r} \left( \frac{N_k(v_j)}{k|H_k|} \right) \right). \quad (8)$$

As shown in Fig. 1, in order to calculate the HOP index of each node pair, a Depth-First Search (DFS)-based is performed to acquire the desired node pairs. In specific, the first-order node pairs are identified by setting the searching depth at one, and the second-order node pairs are identified by setting the depth at two, etc. With the achieved node pairs, we further calculate the HOP index between an arbitrary pair of nodes with (8). It should be pointed out that we can set $r$ equal to the diameter [30] of a target graph to ensure that the acquired HOP indices cover every possible node pair, design inevitably results in heavy computation and storage burden. Therefore, thanks to the well-known *six degrees of separation* [31], we wisely set $r \leq 6$ in our study.

### B. An Iterative Network Reconstruction Scheme

Most existing SNMF-based community detectors are built depending on FOP only, which is inadequate to precisely represent a network due to the following two-fold reasons:
a) The observed edges in a large-scale network are highly limited, leading to great difficulty in achieving its accurate representation; and
b) An SNMF-based community detector is insensitive to the HOP among nodes during its training process.

Nevertheless, we can leverage the HOP metric proposed in the Section III-A to reveal useful HOP information in a network as a highly valuable kind of supplementary information for its representation. From this perspective, we can encode HOP into the original network to obtain an enhanced network whose adjacency is $\tilde{A}=[\tilde{a}_{ij}]$, which enables an NMF-based community detector to directly handle them.

As shown in Fig. 1, based on the achieved HOP indices, the network is further iteratively reconstructed as:

$$\tilde{a}_{ij} = \begin{cases} 1, & a_{ij} = 1; \\ \begin{cases} 1, & P(v_i, v_j) \geq \log \varepsilon, \\ 0, & P(v_i, v_j) < \log \varepsilon, \end{cases} & \text{otherwise,} \end{cases} \quad (9)$$

where $\varepsilon$ is a threshold coefficient deciding whether or not the weight between the node pair $\{v_i, v_j\}$ can be added into $\tilde{A}$ according to their HOP index, and $P(v_i, v_j)$ is the HOP index between nodes $v_i$ and $v_j$.

With (9), the network reconstruction scheme is achieved, and as shown in Fig. 1, it is iteratively conducted to propagate the cumulative effects of previously-added edges. Such a process is repeated until a preset number $d$ is reached to enable an HOP-enhanced network and its adjacency matrix $\tilde{A}$. As the reconstruction process progresses, more and more HOP edges are encoded into a target network. By doing so, the network information is continually enhanced to clarify the community structure. However, if the reconstruction process is performed too many times, the HOP information will be overloaded to blur the community structure. Similar to $r$, we also set $d \leq 6$.

### C. A SGN Algorithm

SNMF learns network representation only depending on a single LF matrix to guarantee its rigorous symmetry, such that its representation ability can be restricted due to the so strong constraint [23-27]. To overcome this issue, in this part, we propose to leverage symmetry-regularizations to make an NMF model own the symmetry of a target undirected network. To do this, we introduce two additional and nonnegative variables $Y$ and $U$ into (2), which are also adopted to fit $\tilde{A}$:

$$J_{SGN} = \frac{1}{4} \left\| XX^T - \tilde{A} \right\|_F^2 + \frac{\vartheta}{2} \left\| YU^T - \tilde{A} \right\|_F^2, \quad (10)$$
$$s.t. \ X \geq 0, \ Y \geq 0, \ U \geq 0;$$

where $\vartheta > 0$ is the coefficient balancing the importance of the involved generalized losses, i.e., $\|XX^T - \tilde{A}\|_F^2$ and $\|YU^T - \tilde{A}\|_F^2$.

Then, two equality-regularization terms between $X$ and $Y/U$ are introduced into (10) for connecting their information:

$$J_{SGN} = \frac{1}{4}\|XX^T - \tilde{A}\|_F^2 + \frac{\vartheta}{2}\|YU^T - \tilde{A}\|_F^2 + \frac{\xi}{2}\|X - Y\|_F^2 + \frac{\zeta}{2}\|X - U\|_F^2, \quad (11)$$

$$s.t.\ X \geq 0,\ Y \geq 0,\ U \geq 0.$$

Note that the equality-regularization terms $\|X-Y\|_F^2$ and $\|X-U\|_F^2$ are adopted to transfer the information learnt by $Y$ and $U$ to $X$, thereby enhancing $X$'s representation to $\tilde{A}$ owing to the enlarged solution space by $Y$ and $U$, and the penalty parameters $\xi > 0$ and $\zeta > 0$ tune the tradeoffs between the generalized loss and equality-regularization. To simplify the hyper-parameter tuning, we set $\xi = \zeta = \vartheta$ in this study. Moreover, to represent the local invariance of the network, graph-regularization is incorporated into (11) to achieve:

$$J_{SGN} = \frac{1}{4}\|XX^T - \tilde{A}\|_F^2 + \frac{\lambda}{2}\text{Tr}(X^T \tilde{L} X) + \frac{\vartheta}{2}\left(\|YU^T - \tilde{A}\|_F^2 + \|X - Y\|_F^2 + \|X - U\|_F^2\right), \quad (12)$$

$$s.t.\ X \geq 0,\ Y \geq 0,\ U \geq 0.$$

where $\text{Tr}(\cdot)$ calculates the trace of an enclosed matrix. $\lambda > 0$ is a coefficient of graph-regularization to adjust its effect. $\tilde{L} = \tilde{D} - \tilde{W}$ is the Laplacian matrix where $\tilde{D}$ is a diagonal degree matrix whose entry is calculated as $\tilde{D}_{ii} = \sum_l \tilde{W}_{il}$ and $\tilde{W}$ is a similarity matrix used to measure the closeness between node pairs. As discussed in [29], $\tilde{W}$ and $\tilde{A}$ are numerically equal in our context.

To achieve $X$, $Y$, and $U$, we need to correctly solve the following optimization problem [32, 34, 35, 43-45]:

$$(X,Y,U) \leftarrow \arg\min_{X,Y,U} J_{SGN},\ s.t.\ X \geq 0, Y \geq 0, U \geq 0. \quad (13)$$

Note that the objective function $J(X, Y, U)$ in (13) is nonconvex over $X$, $Y$ and $U$ together. Therefore, it is unrealistic to expect a learning scheme to achieve the global optimum. In order to find its local optimum, by following [33], we adopt an alternating iterative algorithm that follows the Karush-Kuhn-Tucker (KKT) condition to solve (13), and achieve the following Nonnegative Multiplicative Update (NMU) scheme for SGN:

$$x_{ik} \leftarrow x_{ik}\left(1 - \beta + \beta \frac{\left((1+\lambda)\tilde{A}X + \vartheta Y + \vartheta U\right)_{ik}}{\left(XX^T X + 2\vartheta X + \lambda \tilde{D}X\right)_{ik}}\right), \quad (14a)$$

$$y_{jk} \leftarrow y_{jk} \frac{\left(\tilde{A}U + X\right)_{jk}}{\left(YU^T U + Y\right)_{jk}}, \quad (14b)$$

$$u_{sk} \leftarrow u_{sk} \frac{\left(\tilde{A}^T Y + X\right)_{sk}}{\left(UY^T Y + U\right)_{sk}}. \quad (14c)$$

With the scheme (14), an SGN algorithm is achieved. Note that following [29], we adopt the learning rule (14a) with a linearly adjusted multiplicative term for $X$, where $0 < \beta \leq 1$ is an adjusting coefficient that is suggested to be 0.5 following [21, 29]. An HSGN-based community detector first enhances the network by the HOP-incorporated reconstruction scheme, and then performs community detection via the SGN algorithm.

## IV. PERFORMANCE ANALYSIS

### A. General Settings

Following [8], [21], [29], [36], we adopt two metrics, i.e., Normalized Mutual Information (NMI) and Purity, to evaluate the detecting performance of all tested methods.

TABLE II
DETAILS OF EXPERIMENTAL DATASETS

| Datasets | # Nodes | # Edges | K | Description |
|---|---|---|---|---|
| Amazon | 5,112 | 16,517 | 85 | Amazon product [39] |
| YouTube | 11,144 | 36,186 | 40 | Youtube online [39] |
| Friendster | 11,023 | 280,755 | 13 | Friendster online [39] |
| Orkut | 11,751 | 270,667 | 5 | Orkut online [39] |
| LJ | 11,465 | 400,952 | 27 | LiveJournal online [39] |
| Cora | 2,708 | 5,429 | 7 | LINQS [40] |
| Rugby | 854 | 35,757 | 15 | Insight Resources [41] |
| Olympics | 464 | 7,787 | 28 | Insight Resources [41] |

Eight real-world networks are adopted in our experiments to verify out the proposed method, as summarized in Table II.

We compare the proposed HSGN-based community detector against several benchmark and state-of-the-art methods, i.e., NMF [37], SNMF [29], GNMF [36], GSNMF [8], NSED [38], SymNMF [10], HPNMF [21], LINE [19], PGS [22]. For LINE, we first need to learn the embedding of nodes, and then use the k-means algorithm to cluster the learned node embedding to complete community identification. For HSGN, to specify the effects of its each component, we implement three difference version of this model: 1) **HSGN-I** that only adopts the SGN algorithm for community detection, without considering the HOP-incorporated network reconstruction; 2) **HSGN-II** that adopts the proposed HOP-incorporated network reconstruction algorithm and performs community detection via a GSNMF model; and 3) the complete **HSGN** model in Section III-D.

In order to obtain objective experimental results, we adopt the following commonly accepted settings:

a) Following to [22, 29], in order to eliminate initialization biases, we use the same randomly generated arrays whose entries are randomly generated in the range of (0, 0.5] to initialize all NMF-type community detectors.

b) We set all the involved hyper-parameters with their optimal values for compared models. For LINE, we conduct the experiments with its default settings in official toolkits. For all graph-regularization-incorporated models, i.e., GNMF, GSNMF, HPNMF, PGS and HSGN, we artificially tune the graph-regularization coefficient $\lambda$ in the range of $[10^{-2}, 10^{-1}, 10^0, 10^1, 10^2, 10^3]$. In addition, for HSGN, we select $\vartheta$ in the range of $[2^{-8}, 2^{-5}, 2^{-3}, 2^{-1}, 2^0, 2^1]$, and select $\varepsilon$ in the range of [2, 5, 10, 15, 20].

c) The training process of a tested model terminates if: 1) the difference of the objective value between two consecutive iterations is smaller than a preset threshold, e.g., $10^{-1}$; and 2) the number of iterations reaches a preset one, i.e., 200.

Moreover, we let each separate experiment repeat 10 times with different initial hypotheses to achieve the final results.

### B. Sensitivity Analysis

In this part, we study the effects of hyper-parameters in the following.

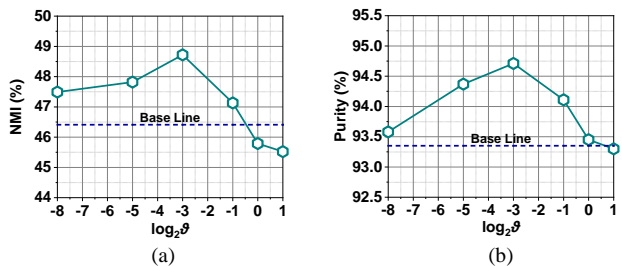

Fig. 2. Effects of $\vartheta$ on Amazon network.

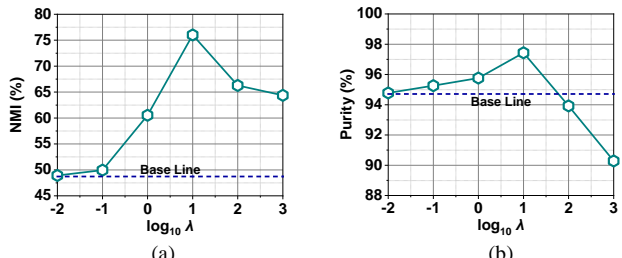

Fig. 3. Effects of $\lambda$ on Amazon network.

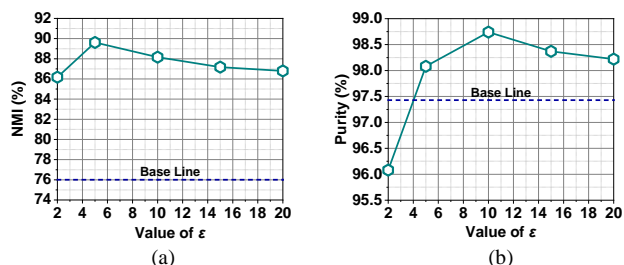

Fig. 4. Effects of $\varepsilon$ on Amazon network.

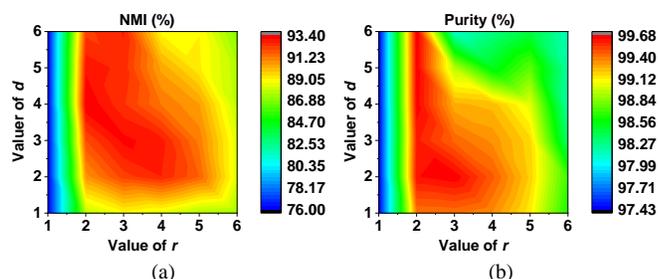

Fig. 5. Effects of $r$ and $d$ on Amazon network.

*1) Effects of $\vartheta$*

We first evaluate the effects of $\vartheta$ by fixing $\lambda=0$ and $\varepsilon \to \infty$. We adopt the results of $\vartheta \to 0$ as the baseline, and report the results on Amazon network in Fig. 2. Similar results can be observed on other networks.

From Fig. 2, we can find that $\vartheta$ has vital effects on HGSN's accuracy. On D1, when $\vartheta \leq 2^{-3}$, HSGN's performance increases as $\vartheta$ increases, and it achieves the highest NMI and Purity as $\vartheta = 2^{-3}$. When $\vartheta$ increases over $2^{-3}$, its NMI and Purity keep on decreasing. When $\vartheta=0$, its NMI and Purity drastically decrease. These results demonstrate that the proposed SGN algorithm plays a vital role in boosting HSGN's accuracy via enlarging its capacity by multiple LF matrices and equality-regularizations.

In following experiments, $\vartheta$ are set in sequence according to the datasets sequence in Table II as $2^{-3}$, $2^{-3}$, $2^{-3}$, $2^0$, $2^{-3}$, $2^0$, $2^1$, and $2^{-1}$, respectively.

*2) Effects of $\lambda$*

We then evaluate the effects of $\lambda$ by fixing $\vartheta$ based on the previous section and $\varepsilon \to \infty$. The tuning range of $\lambda$ is [$10^{-2}$, $10^{-1}$, $10^0$, $10^1$, $10^2$, $10^3$], and the results on Amazon are shown in Fig. 3. Similar situations are encountered on the other networks. Meanwhile, we take the case of $\lambda=0$ as the baseline. In Fig. 3, we see that HSGN's accuracy obviously affected by $\lambda$. On Amazon network, when $\lambda \leq 10$, as $\lambda$ increases, its performance across NMI and Purity keeps on increasing rapidly. Hence, graph-regularization is also very important to make HSGN achieve high detection accuracy.

Note that the selection of $\lambda$ is also dataset-dependent. In our experiments, $\lambda$ is set in sequence according to the datasets sequence in Table II as 10, 10, 0.1, 10, 1.0, 1.0, 10, and 0.1.

*3) Effects of $\varepsilon$*

In this part of experiments, we aim to evaluate the effects of $\varepsilon$. Amazon is used and $\varepsilon$ is tuned in the range of [2, 5, 10, 15, 20]. Note that $\vartheta$ and $\lambda$ are chosen according to the previous sections, and $r$ and $d$ are set at two uniformly. We report the results in Fig. 4, but similar results are encountered on other networks. As shown in Fig. 4, the proposed HOP-incorporated network reconstruction scheme also plays an important role in HSGN's high performance. As depicted in Fig. 4(a), HSGN achieves the highest NMI with $\varepsilon=5$, and as $\varepsilon$ increases over 5, its NMI keeps on decreasing. Taking into account the situations regarding both NMI and Purity, we set $\varepsilon$ at 5 on the Amazon network for HSGN. Further, the parameter $\varepsilon$ is set in sequence according to the datasets sequence in Table II as 5, 5, 2, 2, 5, 2, 2, and 10, respectively.

*4) Effects of $r$ and $d$*

As discussed in Sections III-A and B, based on the six degrees of separation, we set $r \leq 6$ and $d \leq 6$ respectively. Fig. 5 depicts the effects of $r$ and $d$ on Amazon network. Note that similar results are encountered on other networks. According to Fig. 5, we find that HOP enables the iterative reconstruction scheme to improve HSGN's detection accuracy.

As shown in Fig. 5, HSGN's NMI and Purity are always the lowest as $r=1$, where only the FOP information is considered in the reconstruction scheme. However, as $r=2$, we see HSGN's performance watershed. Moreover, as $d>1$, HSGN's detection accuracy can be further boosted. For instance, as depicted in Fig. 5, when $2 \leq r \leq 4$, HSGN's NMI and Purity generally increase as $d$ increases. The results indicate the necessity of the reconstruction scheme's iterative network enhancement based on the incorporated HOP information.

Note that when $r>4$, the performance gain by the network reconstruction scheme gradually vanishes. For instance, as shown in Fig. 5, when $r \geq 4$ and $d \geq 3$, HSGN's NMI and Purity tend to decrease as $d$ increases. Note that the reconstruction scheme tends to incorporate the indirect relationship between two 'long-path-connected' nodes into $\tilde{A}$ as $r$ increases, and it will fill $\tilde{A}$ with more HOP edges as $d$ increases. Both operations actually lead to inappropriate information amplification of the undirected connections among involved nodes. Hence, based on the experiments in this section, we set $r=2$ and $d=3$ for HSGN on all tested network uniformly.

TABLE III
COMMUNITY DETECTION PERFORMANCE (NMI%±STD%) OF TESTED MODELS ON EACH NETWORK.

| Datasets<br>Models | Amazon | YouTube | Friendster | Orkut | LJ | Cora | Rugby | Olympics |
|---|---|---|---|---|---|---|---|---|
| NMF | 42.21±1.46 | 17.26±2.58 | 68.08±6.66 | 29.53±2.13 | 23.47±3.87 | 9.02±3.64 | 35.65±0.40 | 53.30±2.82 |
| SNMF | 45.48±1.72 | 16.95±2.10 | 67.64±5.53 | 29.39±6.36 | 33.78±4.49 | 12.51±0.49 | 34.71±2.03 | 62.44±3.39 |
| GNMF | 65.11±1.43 | 49.45±2.18 | 62.64±6.95 | 59.18±4.02 | 53.94±7.03 | 20.13±0.80 | 44.23±4.00 | 55.08±3.33 |
| GSNMF | 69.41±1.80 | 38.99±2.47 | 45.17±2.89 | 71.92±3.88 | 65.77±1.91 | 18.05±1.66 | 41.32±2.00 | 63.53±1.49 |
| NSED | 46.55±1.87 | 18.44±1.44 | 76.14±5.13 | 31.25±4.93 | 19.79±5.19 | 6.25±3.51 | 34.08±2.27 | 42.32±2.64 |
| SymNMF | 48.30±0.54 | 18.59±1.85 | 78.40±4.49 | 33.75±7.20 | 42.22±2.81 | 15.60±2.52 | 35.40±0.48 | 63.65±0.89 |
| HPNMF | 75.31±2.11 | 50.04±2.54 | 80.55±3.55 | 47.99±8.32 | 68.03±4.51 | 16.40±2.77 | 41.56±2.79 | 64.14±3.43 |
| LINE | 48.23±1.65 | 17.84±1.39 | 70.31±1.95 | 43.31±2.36 | 42.33±1.59 | 4.75±0.16 | 39.66±0.50 | 36.18±0.69 |
| PGS | 84.04±1.36 | 50.63±1.84 | 84.91±3.80 | 76.60±3.19 | 73.44±0.71 | 15.82±3.36 | **49.62±4.05** | 64.51±1.58 |
| HSGN-I | 76.01±1.52 | 50.65±3.80 | 82.98±4.79 | 77.55±7.08 | 71.46±2.28 | 20.61±1.11 | 49.39±6.32 | 68.60±3.60 |
| HSGN-II | 85.39±1.66 | 51.69±1.23 | **93.52±0.11** | 85.44±7.55 | 79.80±3.14 | 22.48±4.92 | 44.00±3.78 | 69.48±4.09 |
| HSGN | **91.94±0.23** | **54.52±0.93** | 89.29±0.23 | **89.12±10.06** | **84.31±0.08** | **29.92±2.02** | 38.57±1.05 | **69.94±1.82** |

TABLE IV
COMMUNITY DETECTION PERFORMANCE (PURITY%±STD%) OF TESTED MODELS ON EACH NETWORK.

| Datasets<br>Models | Amazon | YouTube | Friendster | Orkut | LJ | Cora | Rugby | Olympics |
|---|---|---|---|---|---|---|---|---|
| NMF | 93.30±1.36 | 53.09±1.56 | 81.20±0.02 | 57.20±0.01 | 64.54±0.83 | 48.17±3.43 | 91.52±0.27 | 82.50±1.74 |
| SNMF | 93.56±1.09 | 59.32±7.12 | 78.86±3.19 | 60.86±3.75 | 66.61±5.29 | 53.16±1.50 | 91.50±2.17 | 86.64±1.85 |
| GNMF | 96.75±0.34 | 57.37±1.07 | 89.51±3.30 | 58.19±0.01 | 71.31±1.64 | 58.30±1.12 | 92.93±0.49 | 83.32±1.15 |
| GSNMF | 95.82±1.31 | 57.13±1.21 | 84.74±1.87 | 78.93±0.77 | 70.81±4.10 | 53.42±0.84 | 91.69±2.14 | 87.11±1.57 |
| NSED | 93.71±1.01 | 58.61±4.97 | 85.44±3.76 | 52.27±5.06 | 71.87±2.18 | 46.87±1.21 | 90.33±2.35 | 76.98±1.56 |
| SymNMF | 94.33±0.58 | 58.84±0.16 | 91.82±0.08 | 67.35±8.63 | 71.21±2.84 | 52.86±1.27 | 91.10±0.13 | 86.21±0.61 |
| HPNMF | 94.94±1.18 | 60.10±0.70 | 87.78±2.34 | 65.25±9.92 | 74.67±3.73 | 53.60±1.78 | 94.40±0.33 | 87.46±1.32 |
| LINE | 95.07±1.27 | 58.90±0.43 | 84.97±2.62 | 76.66±3.02 | 71.38±2.18 | 27.81±0.21 | 90.40±1.73 | 26.33±0.99 |
| PGS | 97.86±0.87 | 60.53±0.09 | 89.59±2.97 | 87.11±2.76 | 83.58±1.65 | 60.30±4.71 | 94.36±0.23 | 88.19±0.42 |
| HSGN-I | 97.43±0.89 | 60.48±0.29 | 87.27±2.18 | 77.61±1.24 | 73.45±2.11 | 54.45±2.45 | 94.52±0.11 | 89.31±1.85 |
| HSGN-II | 98.53±0.17 | 60.22±0.53 | **97.44±0.09** | 87.28±4.25 | 84.53±0.33 | 57.34±3.67 | 93.30±1.89 | 90.52±1.83 |
| HSGN | **99.60±0.06** | **60.71±0.14** | 90.13±1.00 | **95.54±5.26** | **89.05±0.76** | **64.45±3.04** | **95.01±0.42** | **90.66±1.64** |

*C. Comparison Results*

We compare HSGN with benchmark and state-of-the-art models to validate its performance. We record the average NMI and Purity values of all compared models on eight networks in Table III and IV. From these results, we conclude that:

a) **HOP-incorporated models, i.e., PGS, HSGN-II, and HSGN, outperform their peers relying on FOP only in terms of community detection accuracy.** From Table III, we see that PGS, HSGN-II and HSGN are outperforms the other models on seven networks out of eight in total, expect that on Rugby the NMI value of HSGN-II is slightly lower than that of GNMF, and the NMI value of HSGN is lower than that of GNMF, GSNMF, LINE and HPNMF. The situation regarding Purity is similar as shown in Table IV. In general, the results tell us that FOP is inadequate and HOP is particularly useful for community detection.

b) **Owing to its iterative network reconstruction scheme and SGN algorithm, HSGN outperforms its peers in achieving highly accurate community detection results.** As shown in Table III, HSGN outperforms its peers in terms of NMI on seven testing cases out of eight in total. Its NMI gain is evident. For instance, compared with the second-best NMI among compared methods, its NMI improvements (i.e., ($NMI_{high}-NMI_{low}$)/$NMI_{low}$) on Amazon, Youtube, Friendster, Orkut, LJ, Cora, and Olympics networks come up to about 9.40%, 7.68%, 5.16%, 16.34%, 14.80%, 48.63%, and 6,76%, respectively. Besides, we can observe the similar conclusion from Table IV.

V. CONCLUSION

A HOP-incorporated, Symmetry and Graph-regularized NMF model for community detection, i.e., HSGN, is presented in this study. It enjoys two-fold virtues: a) A HOP-incorporated iterative network reconstruction scheme to enhance the network information; and b) A SGN algorithm to accurately represent a target network's symmetry by multiple LF matrices standing for a large LF space. Compared with both benchmark and state-of-the-art models, HSGN achieve significantly higher accuracy for community detection. As future work, we plan to address the issue of hyper-parameter adaptation in HSGN through evolutionary computation techniques [42].